\documentclass{ws-procs9x6}

\newcommand{\app}[3]{Astropart.\ Phys.\ {\bf #1} (#2) #3}

\newcommand{\astroph}[2]{{astro-ph/#1}}

\newcommand{\plb}[3]{Phys.\ Lett.\ {\bf B#1} (#2) #3}

\newcommand{\npb}[3]{Nucl.\ Phys.\ {\bf B#1} (#2) #3}

\newcommand{\apj}[3]{Astrophys.\ J.\ {\bf #1} (#2) #3}
\newcommand{\prl}[3]{Phys.\ Rev.\ Lett. {\bf #1} (#2) #3}
\newcommand{\prd}[3]{Phys.\ Rev.\ {\bf D#1} (#2) #3}

\begin{document}

\title{The positron excess and supersymmetric dark matter\footnote{\uppercase{P}reprint: \uppercase{CWRU-P15-02}}}

\author{Edward A. Baltz\footnote{\uppercase{E}-mail: eabaltz@physics.columbia.edu}}

\address{ISCAP, Columbia Astrophysics Laboratory, 550 W 120th St., Mail Code
5247, New York, NY 10027, USA}

\author{\underline{Joakim Edsj\"o}\footnote{\uppercase{R}apporteur,
\uppercase{E}-mail: edsjo@physto.se}}

\address{Department of Physics, Stockholm University, AlbaNova,
SE-106~91~Stockholm, Sweden}

\author{Katherine Freese\footnote{\uppercase{E}-mail: ktfreese@umich.edu}}

\address{Michigan Center for Theoretical Physics, Physics Department,
University of Michigan, Ann Arbor, MI 48109, USA}

\author{Paolo Gondolo\footnote{\uppercase{E}-mail: pxg26@po.cwru.edu}}

\address{Dept.\ of Physics, Case Western Reserve University, 10900 Euclid Ave.,
Cleveland, OH 44106-7079, USA}


\maketitle

\abstracts{Using a new instrument, the HEAT collaboration has confirmed the
excess of cosmic ray positrons that they first detected in 1994.  We explore
the possibility that this excess is due to the annihilation of neutralino dark
matter in the galactic halo. We confirm that neutralino annihilation can
produce enough positrons to make up the measured excess only if there is an
additional enhancement to the signal.  We quantify the `boost factor' that is
required in the signal for various models in the Minimal Supersymmetric
Standard Model parameter space, and find that a boost factor $\geq 30$
provides good fits to the HEAT data.  Such an enhancement in the signal could
arise if we live in a clumpy halo.
%We discuss what part of supersymmetric parameter space is favored (in that it
%gives the largest positron signal), and the consequences for other direct and
%indirect searches of supersymmetric dark matter.
%We also discuss the consequences for other direct and
%indirect searches of supersymmetric dark matter.
}

\section{Introduction}

Several years ago the HEAT collaboration reported an excess of cosmic ray
positrons with energies $\sim$ 10 GeV\cite{heatfrac}.  In 2000 they again
measured this excess using a new instrument, and found excellent
agreement\cite{newdata}.  The possibility that this excess is due to
annihilations of Weakly Interacting Massive Particles (WIMPs) in the galactic
halo, in particular the neutralinos in supersymmetric models, has been
investigated by many authors; the positron fluxes have been calculated
previously\cite{epcont,kamturner,epline,heatprimary,be,ms99}, and attempts to
fit the new HEAT data with positrons from supersymmetric dark matter
annihilations have been made\cite{ep-full,kanepositron,deboer}.

In an earlier work\cite{be} two of us calculated the positron fluxes from
neutralino annihilation in the halo and compared with the excess reported then
by HEAT\cite{heatfrac}.  Recently we investigated the possibility that the HEAT
excess (in light of the new data) could be due to neutralinos\cite{ep-full}.
We will here go through this possibility and comment on some other work along
the same lines.

\section{Supersymmetric model}

We work in the Minimal Supersymmetric Standard Model (MSSM), with the usual
low-energy parameters: the Higgsino mass parameter $\mu$, the gaugino mass
parameter $M_{2}$, the ratio of the Higgs vacuum expectation values $\tan
\beta$, the mass of the $CP$-odd Higgs boson $m_{A}$, the scalar mass parameter
$m_{0}$ and the trilinear soft SUSY-breaking parameters $A_{b}$ and $A_{t}$ for
the third generation.  More details are presented elsewhere\cite{bg,coann}.

We will consider the case when the lightest neutralino is the lightest
supersymmetric particle and we will use {\sffamily DarkSUSY}\cite{DarkSUSY} for
our calculations. We calculate the relic density including coannihilations
between neutralinos and charginos\cite{coann,GondoloGelmini}.  We will here
only include models where $0.05\le\Omega_\chi h^2\le0.25$ ($\Omega_{\chi}$ is
the density in units of the critical density and $h$ is the present Hubble
constant in units of $100$ km s$^{-1}$ Mpc$^{-1}$).
%This region can be narrowed somewhat if we consider the
%results of CMB measurements (summarized in e.g.\ \cite{cmbdata}),
%which favor $\Omega_\chi h^2=0.14\pm0.05$.

We have used the scans over the MSSM parameter space developed in our earlier
work\cite{ep-full}. For each model, we have check if it is excluded by any
accelerator constraints. The most important of these are the LEP
bounds\cite{pdg2000} on the lightest chargino mass (88.4 GeV if $|
m_{\chi_{1}^{+}} - m_{\chi^{0}_{1}} | > 3$ and 67.7 GeV otherwise) and on the
lightest Higgs boson mass $m_{h}$ (which ranges from 91.5--112 GeV depending on
$\tan\beta$) and the constraints from $b \to s \gamma$\cite{cleo} (we use the
LO implementation in DarkSUSY\cite{DarkSUSY}).

\section{Positron spectra and fits to the HEAT data}

We obtain the positron flux from neutralino annihilation in the galactic
halo\cite{be}.  The model is a true diffusion model and assumes that the
diffusion region of tangled galactic magnetic field is an infinite slab.  This
approximation is reasonable since most of the positrons are emitted quite
nearby so that the outer radial boundary is unimportant.  Furthermore, energy
losses due to synchrotron radiation and inverse Compton scattering from the
cosmic microwave background and from starlight are included.  This model
roughly agrees with earlier work\cite{kamturner}, though the inclusion of
inverse Compton scattering from starlight is crucial as it doubles the energy
loss rate.

In calculating the observed positron flux from annihilations in the halo, we
encounter several astrophysical uncertainties.  First, cosmic ray propagation
is not perfectly understood, though the errors are unlikely to be larger than a
factor of two.  More importantly, the structure of the galactic dark halo is
unknown.  Any clumpiness in the halo serves to enhance the signal.  There is no
compelling argument for any particular value of the enhancement factor, be it
unity or in the thousands or more.

We will parameterize the clumpiness as a boost factor, $B_s$, which we define
to be the boost factor that the WIMP annihilation signal from a smooth galactic
halo must be multiplied by to match the HEAT data.  However, we must be careful
that in postulating a boost factor, we do not overproduce the other products of
neutralino annihilation, especially antiprotons and gamma rays\cite{clumpy}.
We do have some freedom here, in that the boost factors for positrons,
antiprotons and gamma rays are not necessarily equal, as their propagation is
not the same. It turns out that for a distribution of clumps that follow that
of the smooth component, the antiproton flux is not boosted as much at the
positron flux\cite{clumpy,ep-full}.

We fit the full positron dataset of the HEAT experiments (1994 and 1995
combined data\cite{heatfrac} and the 2000 data\cite{newdata}).
%We use the
%positron fraction data, as the error bars are smaller and the data cleaner.
%The full dataset consists of twelve independent measurements of the positron
%fraction at various energies.

We will in the following assume that the standard prediction for the positron
background\cite{moskstrong,moskstrongnew} is correct to within a normalization
factor $N$.  We find that the best fit normalization of the background with no
signal from neutralinos is $N=1.14$, with $\chi^2=3.33$ per degree of freedom.
When adding the signal, we make a simultaneous fit of the normalization of the
background $N$ and the normalization of the signal $B_s$, for each
supersymmetric model in our database of models.  We say that a given model
``gives a good fit to the positron data'' when: (1) the background-plus-signal
fit fits the data better than the background-only fit with a decrease in
$\chi^2$ per degree of freedom greater than unity, namely the
background-plus-signal fit has $\chi^2\le2.33$ per degree of freedom; (2) the
best fit normalization of the background $N$ is between 0.5 and 2.0, namely the
background calculation\cite{moskstrong} is correct to within a factor of two
according to the best fit.

%\begin{figure}[t]
%\centerline{\epsfxsize=0.5\textwidth\epsfbox{fig/figure4_color.eps}}   
%\caption{Boost factor versus antiproton flux.  The trend that models with small
%antiproton fluxes require large boost factors in the positron signal reiterates
%the statement that there is a significant correlation between the antiproton
%and positron fluxes.  The hatched band indicates the BESS measurement
%\protect\cite{BESSdata}.}
%\label{fig:pbar}
%\end{figure}

The positron fluxes are more than an order of magnitude smaller than the HEAT
measurements, and we find that the best fit normalizations of the signal $B_s$
lie between 30 and $10^{10}$.  Values of $B_s$ as large as $10^{10}$ are hardly
realistic, but $B_s$ up to 100--1000 might be acceptable given the
uncertainties in the halo structure. For each model we than check that, with
this boost factor $B_s$, it is not excluded by other experiments. As mentioned
earlier, the antiprotons are the most restrictive in this sense and we have
applied the antiproton constraint saying that the standard calculation of the
antiproton flux (boosted with an antiproton boost factor derived from
$B_s$\cite{ep-full}) does not yield more than at most a factor of 5 too high
antiproton fluxes compared to the BESS measurements\cite{BESSdata}. The factor
of 5 is chosen as an upper limit on the propagation uncertainties.
%In Fig.~\ref{fig:pbar} we show the boost factors needed to get good fits to the HEAT data versus the flux of antiprotons. We see that many of the models with low boost factors are excluded by the BESS measurements. There are however models with good fits that are not excluded by the antiproton data, for which we have used the combined 1995 and 1997 data of the BESS collaboration \cite{BESSdata}.

\begin{figure}[t]
\centerline{\epsfxsize=0.49\textwidth\epsfbox{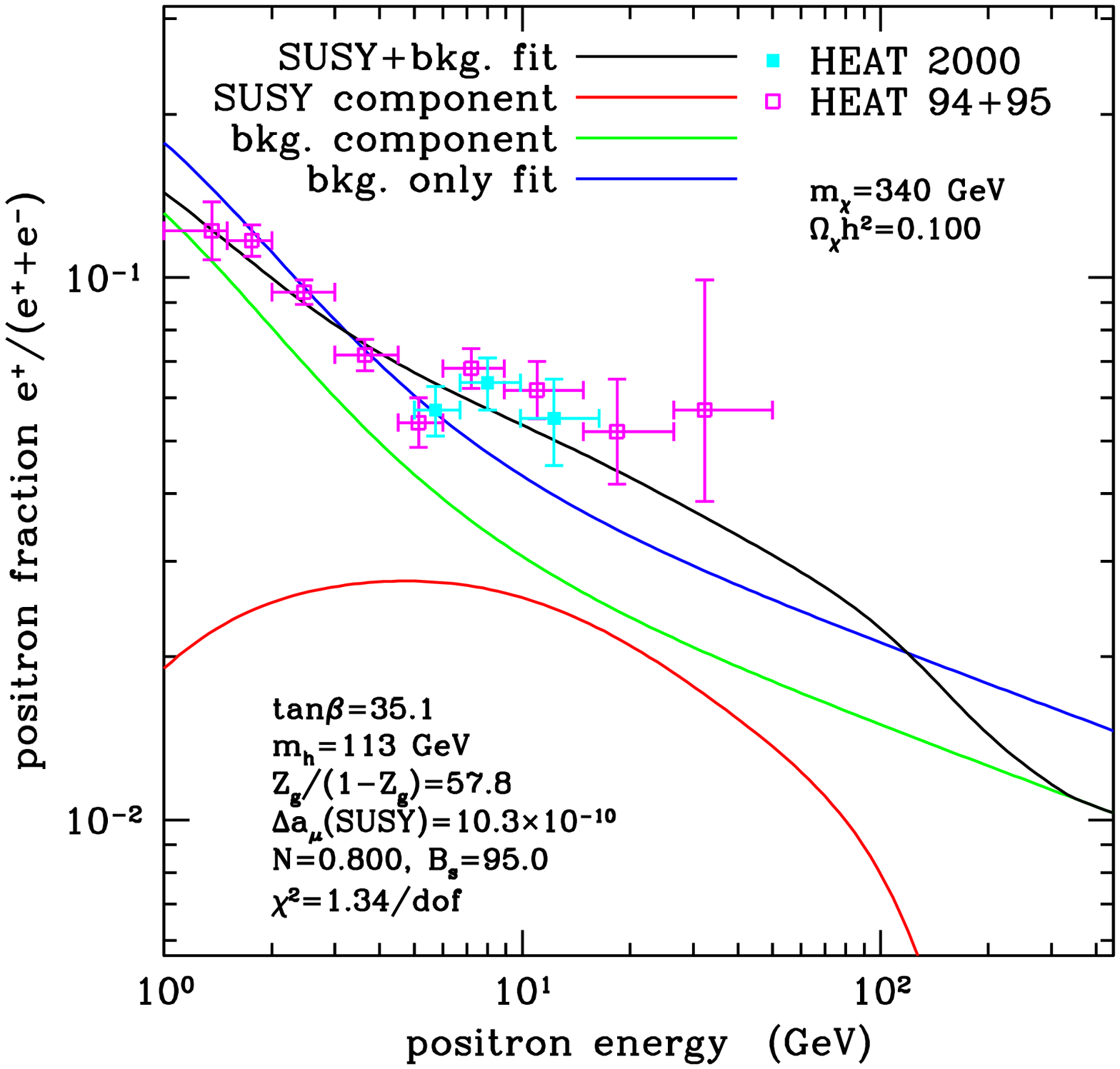}
\epsfxsize=0.49\textwidth\epsfbox{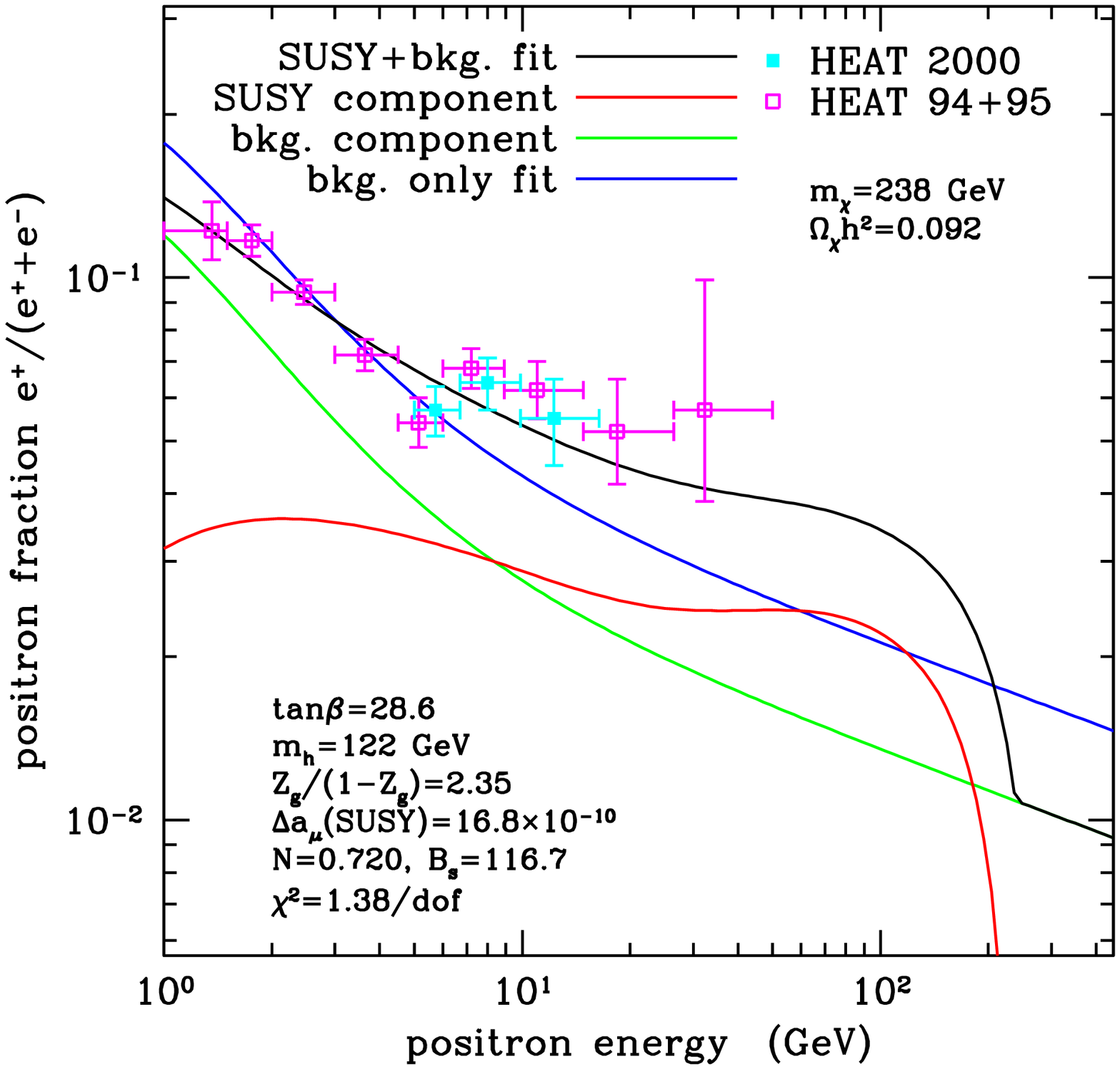}}   
\caption{Positron fraction data and fits.  We illustrate positron data from
HEAT 94+95 and HEAT 2000, a background only fit, and a SUSY+background fit for
two interesting MSSM models.  Two additional curves separately display the SUSY
and background components of the combined SUSY+background fit.  These models
are gaugino dominated and the model in a) has positrons primarily from
hadronization, while the model in b) has hard positrons from direct gauge boson
decays.}
\label{fig:spectrum}
\end{figure}

In Fig.~\ref{fig:spectrum} we plot the positron data from the HEAT 94+95 and
HEAT 2000 experiments, together with the background only fit, and two
interesting SUSY models that have good fits.  The apparent sharp increase in
the positron fraction around 7 GeV is not evident in any of our SUSY models,
even before the smoothing effects of energy loss on the spectrum. In principle
one could envision a sharp edge coming from the direct decay of gauge
bosons. However, we would then get positrons also from the hadronic decay of
the gauge bosons and this would entirely dominate the flux at 7 GeV as seen in
Fig.~\ref{fig:spectrum}b.

Kane et al.\ show that even a monochromatic signal of positrons would not be
able to reproduce the sharp bump seen in the data\cite{kanepositron}.  A
monochromatic signal is smeared by energy losses and does not give the sharp
feature indicated in the data. However, a way to sharpen the bump would be to
have them all come from a nearby clump which is smaller than the propagation
length. Then a line signal would not be smeared out.  This problem has not yet
been treated in depth, but is a bit less appealing since it requires
fine-tuning of both the particle physics and the astrophysics model to explain
the data.

\section{Conclusions}

We have explored models in the Minimal Supersymmetric Standard Model (MSSM)
parameter space to find how large the boost factor must be for each of the
models. The lowest boost factor we found is roughly $30$ for a WIMP that is
primarily a Bino in content with mass $160$ GeV.  For $B_s<100$, we find that
the models are gaugino--dominated, though some have significant Higgsino
fractions.  The masses of the models are in the range 150--400 GeV for the most
part.  For $100<B_s<1000$, the masses are as large as 2 TeV, and some very pure
Higgsinos become allowed.

Fitting the new HEAT data with positrons from supersymmetric dark matter has
been done by other groups\cite{kanepositron,deboer} where they come to more or
less the same conclusions, i.e.\ that large boost factors are needed and that
the fits are better than the background only fits, but they are not excellent.

%The analysis in \cite{deboer} is done in the mSUGRA framework (without taking
%the antiproton constraint into account).

\section*{Acknowledgments}
J.E.~thanks the Swedish Research Council for support.

%%%%%%%%%%


\begin{thebibliography}{0}

\bibitem{heatfrac} S. W. Barwick et al.\ (HEAT Collaboration),
\apj{482}{1997}{L191}.

\bibitem{newdata} S.~Coutu et al.\ (HEAT-pbar Collaboration), in Proceedings of
27th ICRC (2001).

\bibitem{epcont} J.~Silk and M.~Srednicki, \prl{53}{1984}{624}; S.~Rudaz and
  F.~Stecker, \apj{325}{1988}{16}; J.~Ellis, R.A.~Flores, K.~Freese, S.~Ritz,
  D.~Seckel and J.~Silk, \plb{214}{1989}{403}; F.~Stecker and A.~Tylka,
  \apj{336}{1989}{L51}; D.~Eichler \prl{63}{1989}{2440}.

\bibitem{kamturner} M.~Kamionkowski and M.~S.~Turner, \prd{43}{1991}{1774}.

\bibitem{epline} M.S.~Turner and F.~Wilczek, \prd{42}{1990}{1001}; A.J.~Tylka,
\prl{63}{1989}{840}.

\bibitem{heatprimary} S. W. Barwick et al.\ (HEAT Collaboration),
\apj{498}{1998}{779}; S.~Coutu et al.\ (HEAT Collaboration),
\app{11}{1999}{429}.

\bibitem{be} E. A. Baltz and J. Edsj\"o, \prd{59}{1999}{023511}.

\bibitem{ms99} I.~V.~Moskalenko and A.~W.~Strong, \prd{60}{1999}{063003}.

\bibitem{ep-full}
E.A.~Baltz, J.~Edsj\"o, K.~Freese and P.~Gondolo,
Phys.\ Rev.\ {\bfseries D65} (2001) 063511.

\bibitem{kanepositron} G.L.~Kane, L.-T.~Wang and J.~D.~Wells, 
Phys.\ Rev.\ {\bfseries D65} (2002) 057701;
%hep-ph/0108138 (2001);
G.L.~Kane, L.-T.~Wang and T.T.~Wang, Phys.\ Lett.\ {\bfseries B536} (2002) 263.
%hep-ph/0202156.

\bibitem{deboer} W.~de Boer, C.~Sander, M.~Horn and D.~Kazakov, astro-ph/0207557.

\bibitem{bg} L.~Bergstr\"om and P.~Gondolo, \app{5}{1996}{263}.

\bibitem{coann} J.~Edsj{\"o} and P.~Gondolo, \prd{56}{1997}{1879}.

\bibitem{DarkSUSY} P. Gondolo, J. Edsj\"o, L. Bergstr\"om, P. Ullio, and E.A.
Baltz, \astroph{0012234}{2000}.

\bibitem{GondoloGelmini}
P.~Gondolo and G.~Gelmini, \npb{360}{1991}{145}.

\bibitem{cmbdata}
X.~Wang, M.~Tegmark and M.~Zaldarriaga, 
Phys.\ Rev.\ {\bfseries D65} (2002) 123001.
%\astroph{0105091}{2001}.

\bibitem{pdg2000}
D.~E.~Groom et al. (Particle Data Group), Eur.~Phys.~J.~C, {\bf 15}, 1 (2000).

\bibitem{cleo} M.~S.~Alam et al.\ ({\sc Cleo} Collaboration),
\prl{71}{1993}{674} and \prl{74}{1995}{2885}.

\bibitem{clumpy} L. Bergstr\"om, J. Edsj\"o, P. Gondolo, and P. Ullio,
\prd{59}{1999}{043506}.

\bibitem{moskstrong} I.~V.~Moskalenko and A.~W.~Strong, \apj{493}{1998}{694}.

\bibitem{moskstrongnew} I.~V.~Moskalenko, A.~W.~Strong, J.~F.~Ormes and
M.~S.~Potgieter, \apj{565}{2002}{280}.
%astro-ph/0106567

\bibitem{BESSdata} S.~Orito et al., \prl{84}{2000}{1078}.

\end{thebibliography}
\end{document}